\documentclass[amsmath,amssymb]{article}
\def\@preprint{}
\newcommand\preprint[1]{\@preprint{\hfill #1}}

\newcommand{\maketitleNew}[1]{
	
	\setcounter{page}{0}
	\noindent{\small\scshape}\@preprint\par
	\preprint{MITP/21-058, LTH 1283}
	{\let\newpage\relax\maketitle}
	\maketitle
	\thispagestyle{empty}
}
\pdfoutput=1
\usepackage{authblk}
\usepackage{abstract}
\usepackage[footnotesize]{caption}
\usepackage{subcaption}
\usepackage{cite}
\usepackage{amsmath}
\usepackage{amssymb}
\usepackage{graphicx} 
\usepackage[latin1]{inputenc}
\usepackage{mathrsfs}
\usepackage[dvipsnames]{xcolor}
\usepackage{hyperref}
\usepackage[hmarginratio=1:1,top=32mm,columnsep=20pt]{geometry}
\usepackage{paralist}
\usepackage{multirow}
\usepackage{multicol}
\usepackage{blindtext}
\usepackage{upgreek}
\usepackage{soul}

\newcommand{\bea}{\begin{eqnarray}}
\newcommand{\eea}{\end{eqnarray}}
\newcommand{\be}{\begin{equation}}
\newcommand{\ee}{\end{equation}}
\newcommand{\ba}{\begin{array}}
	\newcommand{\ea}{\end{array}}

\def\gsim{\mathrel{\rlap{\lower4pt\hbox{\hskip1pt$\sim$}}
		\raise1pt\hbox{$>$}}}

\renewcommand\Affilfont{\normalsize\itshape}

\title{\fontsize{14pt}{10pt}\selectfont 
	\textbf{
		Explaining excesses in four-leptons at the LHC with a double peak from a CP violating Two Higgs Doublet Model
	}
}
\author[1]{\normalsize Stefan~Antusch\thanks{\texttt{stefan.antusch@unibas.ch}}}
\author[2]{\normalsize Oliver~Fischer\thanks{\texttt{oliver.fischer@liverpool.ac.uk}}}
\author[3]{\normalsize A.~Hammad\thanks{\texttt{ahmed.hammad@unibas.ch}}}
\author[4]{\normalsize Christiane~Scherb\thanks{\texttt{cscherb@uni-mainz.de}}}

\affil[1]{\Affilfont Department of Physics, University of Basel, \authorcr 
	\Affilfont Klingelbergstr.\ 82, CH-4056 Basel, Switzerland
	\authorcr\mbox{}}
\affil[2]{\Affilfont Department of Mathematical Sciences, University of Liverpool
	\authorcr 
	\Affilfont  Liverpool, L69 7ZL, UK
	\authorcr\mbox{}} 	

\affil[3]{\Affilfont Institute of Convergence Fundamental Studies,
	\authorcr 
	\Affilfont  Seoul National University of Science and Technology,  Seoul, 01811, Korea
	\authorcr\mbox{}}	

\affil[3]{\Affilfont Centre for Theoretical Physics,
	\authorcr 
	\Affilfont  the British University in Egypt, P.O. Box 43, Cairo 11837, Egypt
	\authorcr\mbox{}}	

\affil[4]{\Affilfont PRISMA$^+$ Cluster of Excellence $\&$ Mainz Institute for Theoretical Physics,
	\authorcr 
	\Affilfont  Johannes Gutenberg University, 55099 Mainz, Germany
	\authorcr\mbox{}}

\date{}

\begin{document}
	\maketitleNew
	
	\begin{abstract}
		\noindent 
	
\noindent	
Extended scalar sectors with additional degrees of freedom appear in many scenarios beyond the Standard Model. Heavy scalar resonances that interact with the neutral current could be discovered via broad resonances in the tails of the four-lepton invariant mass spectrum, where the Standard Model background is small and well understood. In this article we consider a recent ATLAS measurement of four-lepton final states, where the data is in excess over the background for invariant masses above 500 GeV. We discuss the possibility that this excess could be interpreted as a ``double peak'' from the two extra heavy neutral scalars of a  CP violating Two Higgs Doublet Model, both coupling to the $Z$ boson. We apply an iterative fitting procedure to find viable model parameters that can match the excess, resulting in a benchmark point where  the observed four-lepton invariant mass spectrum can be explained by two scalar particles $H_2$ and $H_3$, with masses of 540 GeV and 631 GeV, respectively, being  admixtures of the CP eigenstates. Our explanation predicts additional production processes for $t\bar t$, $W^+W^-$, $4b$ and $\gamma\gamma$, some of which have cross sections close to the current experimental limits. Our results further imply that the electric dipole moment of the electron should be close to the present bounds.
\end{abstract}

\newpage

\section{Introduction}
The discovery of the scalar resonance with a mass of about 125 GeV, compatible with the predicted Higgs boson \cite{Khachatryan:2016vau}, completes the Standard Model (SM). 
While currently most searches for physics beyond the SM do not point at new physics beyond the SM, excesses in final states with leptons \cite{Abdallah:2014fra,vonBuddenbrock:2019ajh} and di-photons \cite{Fox:2017uwr,Haisch:2017gql,Biekotter:2020cjs} indicate the possible existence of additional scalar degrees of freedom.
The four-lepton final states are often  referred to as the `golden channel' when searching for heavy scalar resonances, due to small and controllable SM backgrounds. 
The four-lepton analyses from ATLAS \cite{ATLAS:2017tlw} and CMS \cite{CMS:2018mmw,CMS:2017jkd} show enhanced event rates in final states with high invariant mass. These analyses were combined in ref.~\cite{Cea:2018tmm} to demonstrate the compatibility of the data with a broad resonance structure around 700 GeV.
This is compatible with a very recent interpretation of ATLAS data as a second Higgs excitation at 680~GeV \cite{Cosmai:2021hvc}.

Searches for heavy scalars above 600 GeV in $\gamma\gamma$, $Z\gamma$, $ZZ \to 4\ell$, top quarks, pairs of Higgs and $W$ bosons were reviewed and discussed in ref.~\cite{Richard:2020jfd}, supporting the claim made in ref.~\cite{Cea:2018tmm} of a resonance in $ZZ$ around 700 GeV.
This possible resonant enhancement is visible both in gluon and vector boson fusion channels as reported by ATLAS in ref.~\cite{ATLAS:2017tlw}
The more recent analysis of four-lepton final states by the ATLAS collaboration also shows an enhancement of event rates with invariant masses above about 500~GeV \cite{Aad:2021ebo}. This observation was considered in ref.~\cite{Richard:2021edf} to corroborate the possible resonance around 700~GeV. 
The theoretical framework was the Georgi-Machacek model and a cross section for the heavy resonance was found to be $\sim 160$ fb.
It is interesting to note that a broader enhancement of the $b\bar bb\bar b$ final state with invariant mass above 500~GeV is visible, which might be compatible with the observed $4\ell$ excess from ref.~\cite{ATLAS:2020jgy}. On the other hand, a recent analysis searching for heavy diboson resonances in semi-leptonic final states 
is compatible with the SM prediction \cite{ATLAS:2020fry}, however here the backgrounds are at a much higher level and might cover up a possible enhancement.

In this article we consider an explanation of the four-lepton excess at invariant masses above 500 GeV by a ``double peak'' from the two extra heavy neutral scalars of a  CP violating Two Higgs Doublet Model (THDM), where the latter extends the scalar sector of the SM by an additional scalar $SU(2)_L$ doublet field \cite{Haber:1978jt}. Due to CP violation, both extra heavy neutral scalars can couple to the $Z$ boson, thus showing up in the four-lepton invariant mass spectrum. Recently, we studied a class of THDMs with CP violation in ref.~\cite{Antusch:2020ngh}, exploring the testability of CP violation at the LHC and evaluating the current constraints on the model parameters, pointing out the importance of the the four-lepton final state as ``discovery channel''. We will make use of the results of \cite{Antusch:2020ngh} to find a viable set of model parameters that can match the current four-lepton excess. 
The article is structured as follows: in section 2 we briefly review the model framework, in section 3 we describe our analysis and discuss the results, and in section 4 we conclude.

\section{The model}
The THDM was first discussed in ref.~\cite{Lee:1973iz} to discuss the phenomenon of CP violation in the scalar sector. For a comprehensive review we refer the reader e.g.\ to ref.~\cite{Branco:2011iw}. In THDMs, the scalar sector contains two $SU(2)_L$-doublet fields, $\phi_1$ and $\phi_2$, with identical quantum numbers under the SM gauge symmetry group:
\begin{equation}
	\phi_1 = \begin{pmatrix}
	\eta_1^+ \\
	(v_1 + h_1 + i h_3)/\sqrt{2} \\
	\end{pmatrix}
	\hspace{0.5cm} \mbox{and}\hspace{0.5cm}
	\phi_2 = \begin{pmatrix}
	\eta_2^+ \\
	(v_2 + h_2 + i h_4)/\sqrt{2} \\
	\end{pmatrix}\,.
\end{equation}
The components $h_i,\,i=1,...,4$, are real neutral fields, $\eta_i^+,\,i=1,2$ are complex charged fields, and $v_i,\,i=1,2$ are the vacuum expectation values (vevs).
The most general Lagrangian density for the model can be decomposed as 
\begin{equation}
	\mathcal L_{\text{THDM}} = \mathcal L_{\text{SM}, \text{kin}} +  \mathcal L_{\phi, \text{kin}} + V_{\phi} + Y_\phi \,,
\end{equation}
where $ \mathcal L_{\text{SM}, \text{kin}} $ denotes the kinetic terms for SM gauge fields and fermions, $\mathcal L_{\phi, \text{kin}}$ denotes the kinetic terms for the two scalar fields $\phi_i,\,i=1,2$, $V_\phi$ denotes the scalar potential, and $Y_\phi$ contains the Yukawa terms that give rise to the couplings between the SM fermions and the scalar fields. 

Since scalar decays into four-lepton final states (henceforth referred to as $4\ell$) come about from scalar decays into two $Z$ bosons that in turn decay into leptons, we are interested in the interaction of the scalar fields with the neutral current.
There are only two interaction eigenstates that can decay into two $Z$ bosons. 
However, the most general (CP violating) form of the THDM leads to scalar mixing, 
and in the presence of CP violation all the three neutral mass eigenstates $H_i$ of the THDM can mediate the process $pp \to H_i \to ZZ \to 4\ell$, with $i=1,2,3$.

The final state from the process $pp \to H_i \to ZZ \to 4\ell$, with $4\ell = \ell_\alpha^+\ell_\alpha^- \ell_\beta^+\ell_\beta^-$ (and where the considered lepton flavors are $\ell_{\alpha,\beta}^\pm = e^\pm,\,\mu^\pm$) features an invariant mass that reflects the mass of the mediating $H_i$.
The mass eigenstate $H_1$ corresponds to the SM-like Higgs boson with $m_{H_1} \simeq 125$ GeV, whose four-lepton signal has been studied (see examples for ATLAS \cite{ATLAS:2020rej} and CMS \cite{CMS:2021ugl} analyses) and $H_2$ and $H_3$ are assumed to be heavier. As discussed above, in a CP violating THDM the extra neutral scalars $H_2$ and $H_3$ give rise to two additional broad peaks in the $4\ell$ invariant mass spectrum - or to a broad ``double peak'' if the masses are not too separated. This is a feature that is usually not considered when fitting the THDM to the data.

\section{Analysis and results}
We consider the measurements of differential cross-sections in $4\ell$ events in the 139 fb$^{-1}$ data set by the ATLAS collaboration \cite{Aad:2021ebo}.
From the results in the ATLAS publication we use the $4\ell$ differential cross sections, and in particular the invariant mass spectrum ($M_{4 \ell}$).
We digitise the observed event rates, their errors, and the theory prediction.

We use the nine bins from $M_{4\ell}$ between 500 GeV and 900 GeV, six of which show event counts in excess of the theory prediction.
We create a sample of excess events by subtracting the theory prediction from the observed event rates (which means that ``excess events'' per bin can also be negative).

\subsection{Numerical setup}
As an explicit example we consider the same class of THDMs with CP violation and type I Yukawa structure as in ref.~\cite{Antusch:2020ngh}, to which we refer the reader for model details and notation.
We calculate testable properties of the THDM via numerical tools that include the following recent constraints:\footnote{We use the same limits as detailed in sec.\ 3 of ref.~\cite{Antusch:2020ngh}.}
$B$ physics data using FlavorKit~\cite{Porod:2014xia}, Higgs data using HiggsBounds~\cite{Bechtle:2013wla} and HiggsSignals~\cite{Bechtle:2013xfa}, and electric dipole moments with the formulae from refs.~\cite{Abe:2013qla,Chun:2019oix}.

The process $pp\to H_i \to 4\ell$ for the heavy scalars with indices $i=2,3$ is calculated in MadGraph~\cite{Alwall:2014hca}, including the effective gluon-Higgs vertex via SPheno~\cite{Porod:2003um, Porod:2011nf} and QCD corrections~\cite{Belyaev:2012qa}.
We note at this point that the inclusive $4\ell$ invariant mass spectrum is simulated to include the interference between the scalars.
A fast detector simulation is done with 500k events per sample with Delphes \cite{deFavereau:2013fsa} using the standard ATLAS detector card.
From the reconstructed events we read out the invariant mass spectra.

\subsection{First approximation}
We searched for points with masses $m_{H_2}$ and $m_{H_3}$ around 500 GeV and 700 GeV, respectively, and with total cross sections for the $4\ell$ final state that are of similar magnitude.
We found the benchmark point $P_1$ with $m_{H_2}^{P_1}=535$ GeV and $m_{H_3}^{P_1}=703$ GeV and total $4\ell$ cross sections $\sigma_{H_2\to 4\ell}^{P_1} = 1.3$ fb and $\sigma_{H_3\to 4\ell}^{P_1} = 0.86$ fb.
For $H_2$ and $H_3$ the signal selection efficiency based on the experimental selection criteria is found to be $\epsilon_{4\ell} \sim 0.3$ and not too dependent on the scalar mass.
We simulated exclusively $pp \to H_2 \to 4\ell$ and $pp \to H_3 \to 4\ell$ from which we obtained the invariant mass spectra $\rho_2$ and $\rho_3$, respectively.
Notice that at this step, the interference between the scalars is not taken into account.

Next we performed a simple $\chi^2$ analysis.
We varied the signal peaks of the two spectra with the two parameters $\delta m_j$, such that the masses are given by $m_{H_j}=m_{H_j}^{P_1}+\delta m_j$.
We also introduce the signal strength multipliers $s_j$ which are multiplied with the invariant mass spectra $\rho_{H_j}$.
We construct the $\chi^2$:
\begin{equation}
\chi^2_{\rm sig}(\delta m_2,\delta m_3,s_2,s_3) = \sum_i \frac{(b_{sig,i}(\delta m_2,\delta m_3,s_2,s_3)-b_i)^2}{\delta_{obs,i}^2 + \delta_{sys,i}^2}
\label{eq:chi-squared}
\end{equation}
where $i$ is the bin number, $b_i$ is the measured event rate $\delta_{obs,i} = \sqrt{b_i}$, $\delta_{syst,i}=10\%$. The signal rate for bin $i$ is
\begin{equation}
b_{sig,i}(\delta m_2,\delta m_3,s_2,s_3) = b_{SM,i} + \int_i (s_2 \cdot \rho_{H_2}(E-\delta m_2)+ s_3\cdot\rho_{H_3}(E-\delta m_3)) dE\,,
\label{eq:partial-spectra}
\end{equation}
with $b_{SM,i}$ being the SM prediction, $\rho_{H_j}$ the signal distributions, and the parameters $\delta m_j$ and $s_j$ are varied to minimise the $\chi^2$.

\subsection{Iterative analysis}
Notice that the signal rate as described in \eqref{eq:partial-spectra} distorts the invariant mass spectrum and thus disconnects it from the underlying benchmark point.
However, the distorted spectrum can be used to locate the masses and event rates that are preferred by the fit to the data.

Consequently we use the best-fit values for the masses and the total $4\ell$ cross sections (converted from the fiducial cross sections using the signal selection efficiency and the integrated luminosity) as selection criteria to find a new benchmark point.
From a fine grained scan in the model parameters we select a point $P_{n}$ that has masses $m_{H_j}$ and cross sections $\sigma_{H_j \to 4\ell}$ that are as close to the best-fit results for the masses and event rates of the previous point $P_{n-1}$ as possible.

For each new point $P_{n}$ we create an inclusive $4\ell$ invariant mass spectrum $\rho_{incl}^{P_{n}}$ with two peaks around $m_{H_2}$ and $m_{H_3}$, including the interference between $H_2$ and $H_3$. We remark that the interference with $H_1$ is negligible for the here relevant mass scales of $H_2$ and $H_3$, which we verified by computation.
We separate the spectrum into $\rho_{H_2}^{P_{n}}$ and $\rho_{H_3}^{P_{n}}$ at the minimum between the two peaks and fit the parameters $\delta m_j$ and $s_j$ via the two partial spectra as in eq.~\eqref{eq:chi-squared} to the data.

Once we have a benchmark point with a spectrum that provides a good fit, we carry out a Bayesian fit of the parameters $\delta m_j,\,j=2,3$ and $s_j,\,j=2,3$ to establish the Bayesian confidence limits on the parameters.

\subsection{Results}
Our analysis converged sufficiently after six iterations. 
We perform an even more fine-grained parameter space scan and select parameter space points with $m_{H_2}, m_{H_3}$, $\sigma_{H_2\to 4\ell}$ and $\sigma_{H_3\to 4\ell}$ within the 90\% Bayesian confidence interval around the last iteration's best fit parameters, which are:
\begin{equation}
\begin{array}{cc}
521.1~\text{GeV} \leq m_{H_2} \leq 562.9~\text{GeV}, & \qquad 0.3~\text{fb} \leq \sigma_{H_2\to 4\ell} \leq 0.6~\text{fb}, \\
602.2~\text{GeV} \leq m_{H_3} \leq 655.9~\text{GeV}, & \qquad 0.3~\text{fb}\leq \sigma_{H_3\to 4\ell} \leq 0.6~\text{fb}.
\end{array}
\end{equation}

\begin{figure}
\centering
\includegraphics[width=0.5\textwidth]{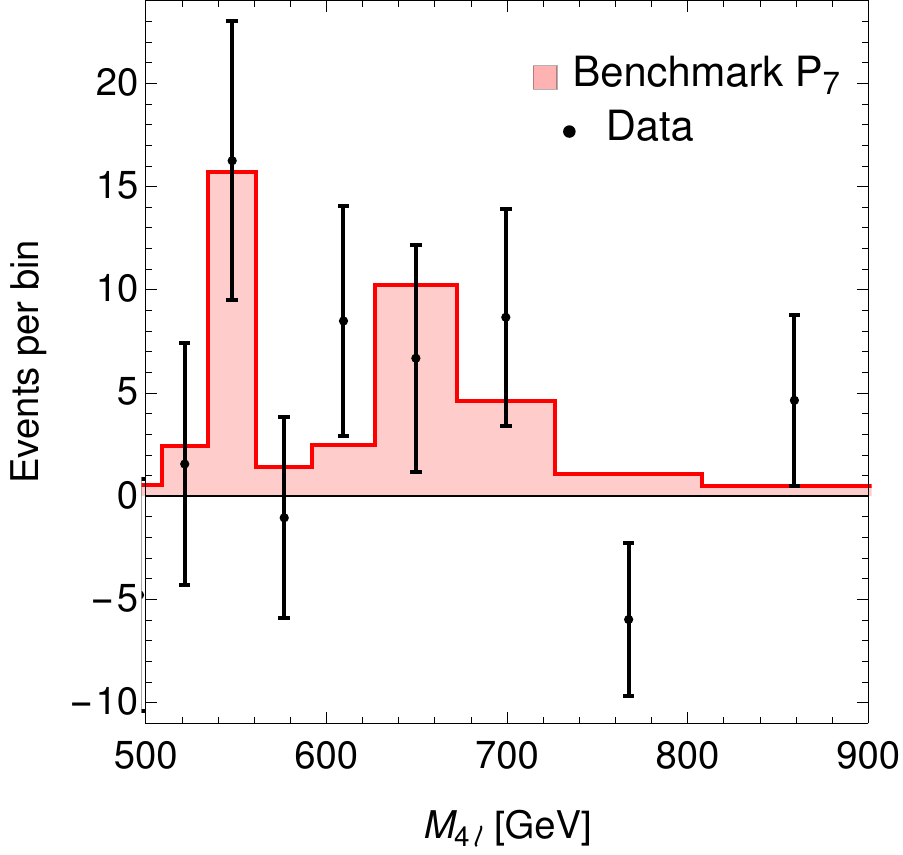}

\caption{Invariant mass spectrum of the best-fit points for the ``best-fit benchmark point'' $P_7$, for details, see text. The black dots with error bars denote the difference between observed and predicted data from the four-lepton invariant mass spectrum in ref.~\cite{Aad:2021ebo}.
}
\label{fig:bestfitpoint}
\end{figure}

In this scan we find the benchmark point $P_7$ that is defined by the following model parameters: $\tan\beta = 20$, $\eta = 0.659$, $\lambda_1 = 0.7$, $\lambda_2 = 0.099$, $\lambda_3 = 4.97$, $\lambda_4 = 6.45$ and $\Re(\lambda_5) = 1.792$.
These parameters give rise to $m_{H_2}^{P_7}=540$~GeV, $m_{H_3}^{P_7}=631$ GeV, and $\sigma_{H_2\to 4\ell}^{P_7} = 0.45$ fb, $\sigma_{H_3\to 4\ell}^{P_7} = 0.42$ fb. 
We call this point the ``best-fit benchmark point'' as is provides a very good fit to the spectrum with a $\chi^2=5.76$, which is to be compared to the SM value for all bins above 500 GeV. For the SM we get a $\chi^2_{SM} = 21.0~(16.9)$ corresponding to an upward fluctuation with a p-value of 0.007 (0.03) considering statistical errors only (all errors).
The contribution of the two neutral scalar particles to the four-lepton invariant mass spectrum for the ``best fit benchmark point'' $P_7$ is shown in fig.~\ref{fig:bestfitpoint}.
The striking feature of the spectrum is the wide range of $M_{4\ell}$ that receives contributions from the two heavy scalars.

\begin{figure}

\centering
Top: $WW$

\begin{minipage}{0.49\textwidth}
	\includegraphics[width=\textwidth]{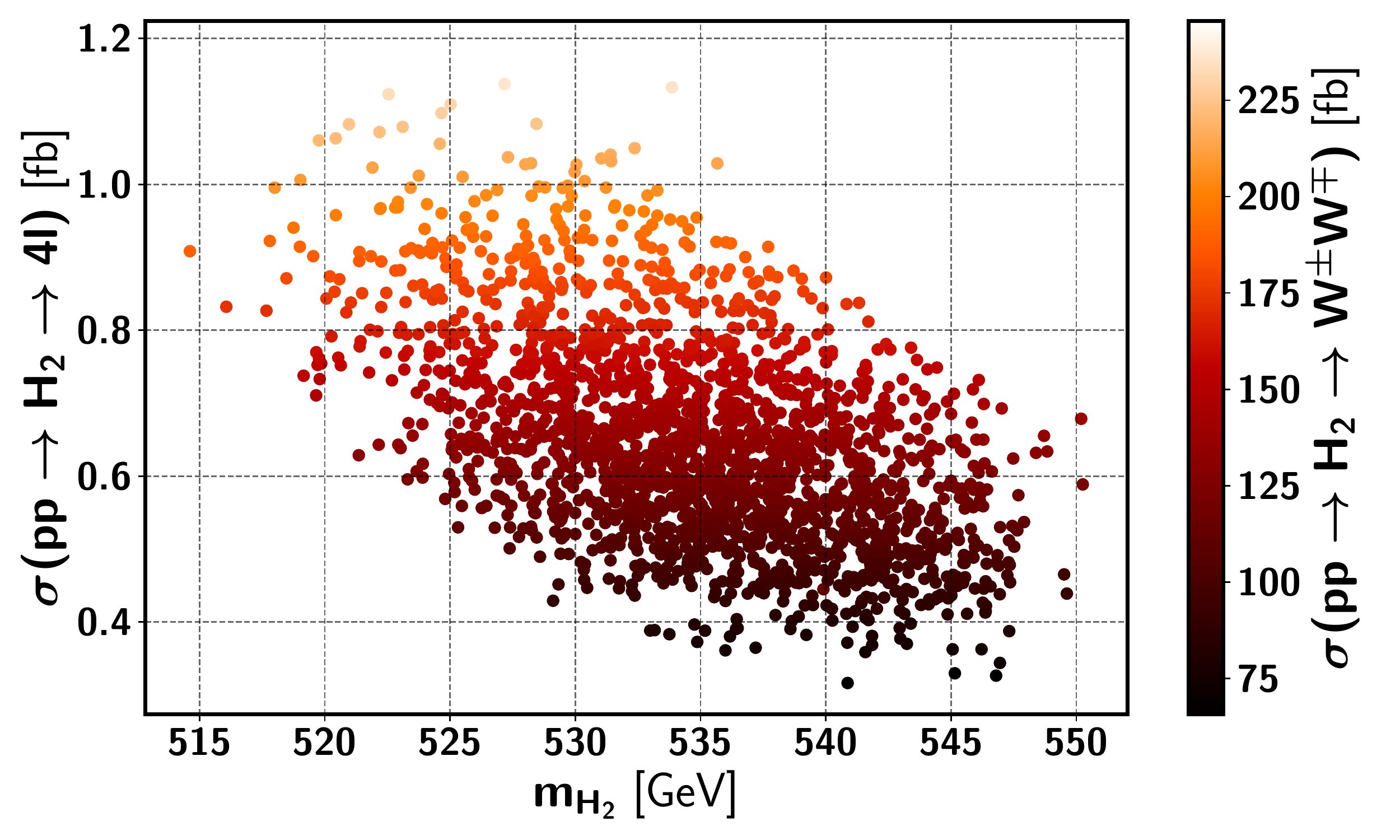}
\end{minipage}
\begin{minipage}{0.49\textwidth}
	\includegraphics[width=\textwidth]{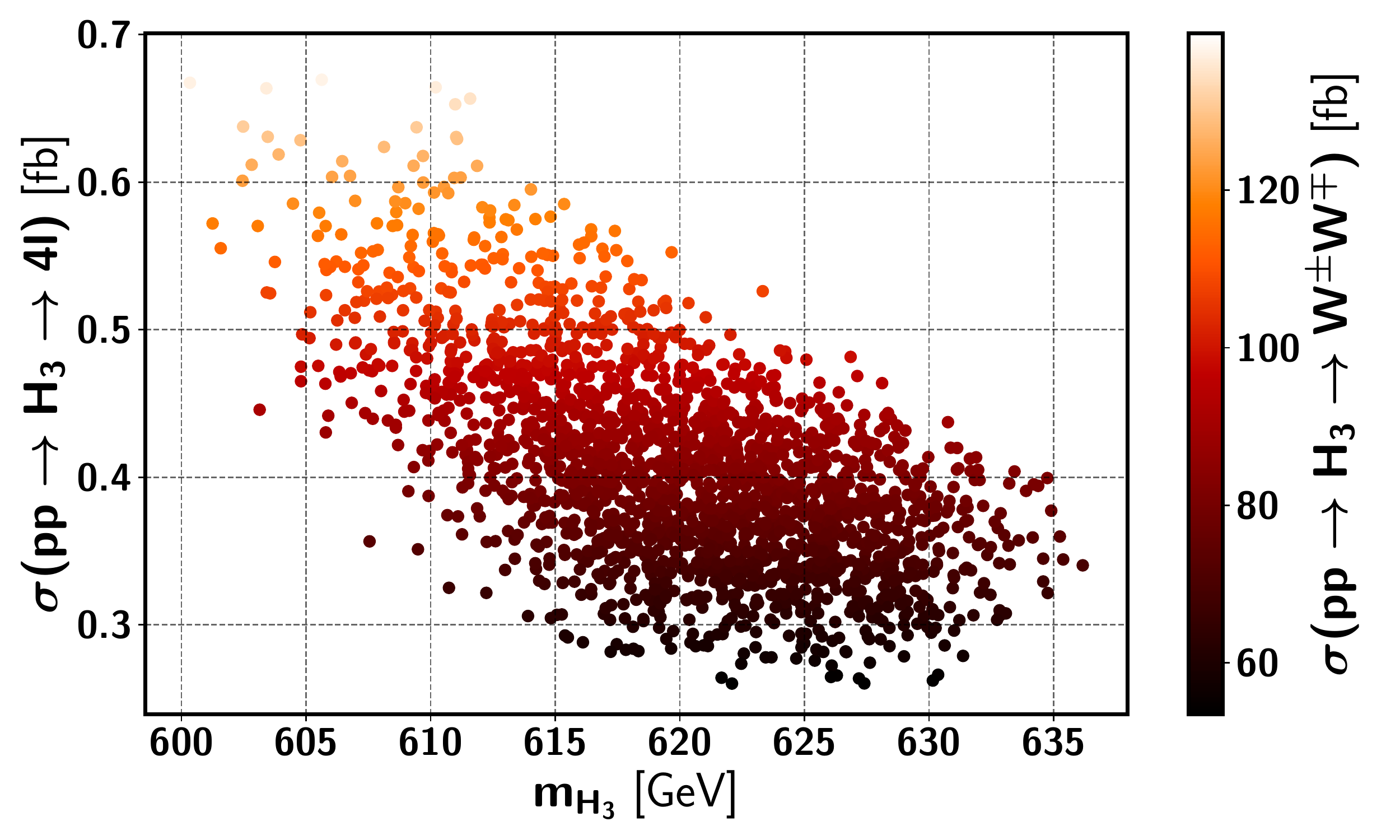}
\end{minipage}

Middle: $t\bar t$

\begin{minipage}{0.49\textwidth}
	\includegraphics[width=\textwidth]{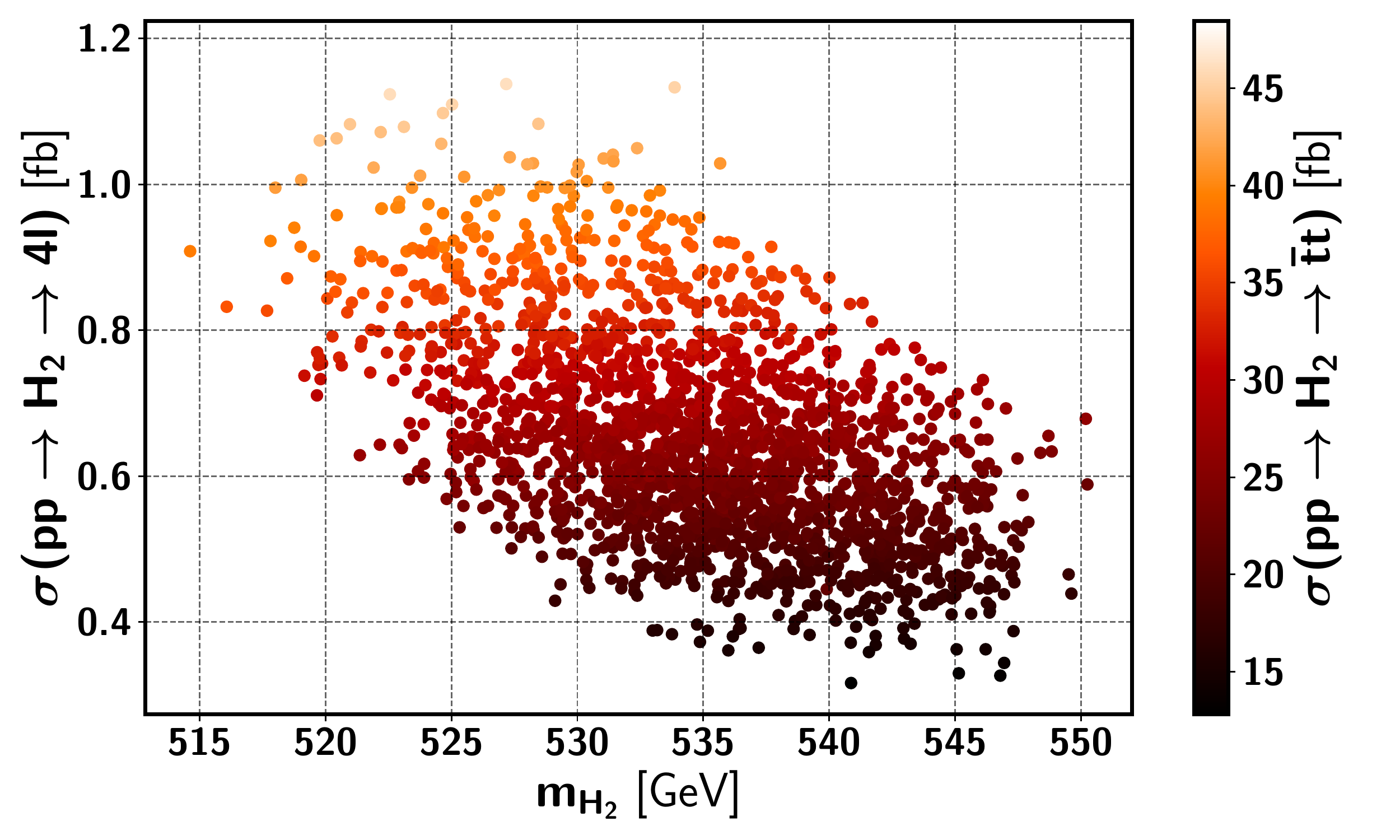}
\end{minipage}
\begin{minipage}{0.49\textwidth}
	\includegraphics[width=\textwidth]{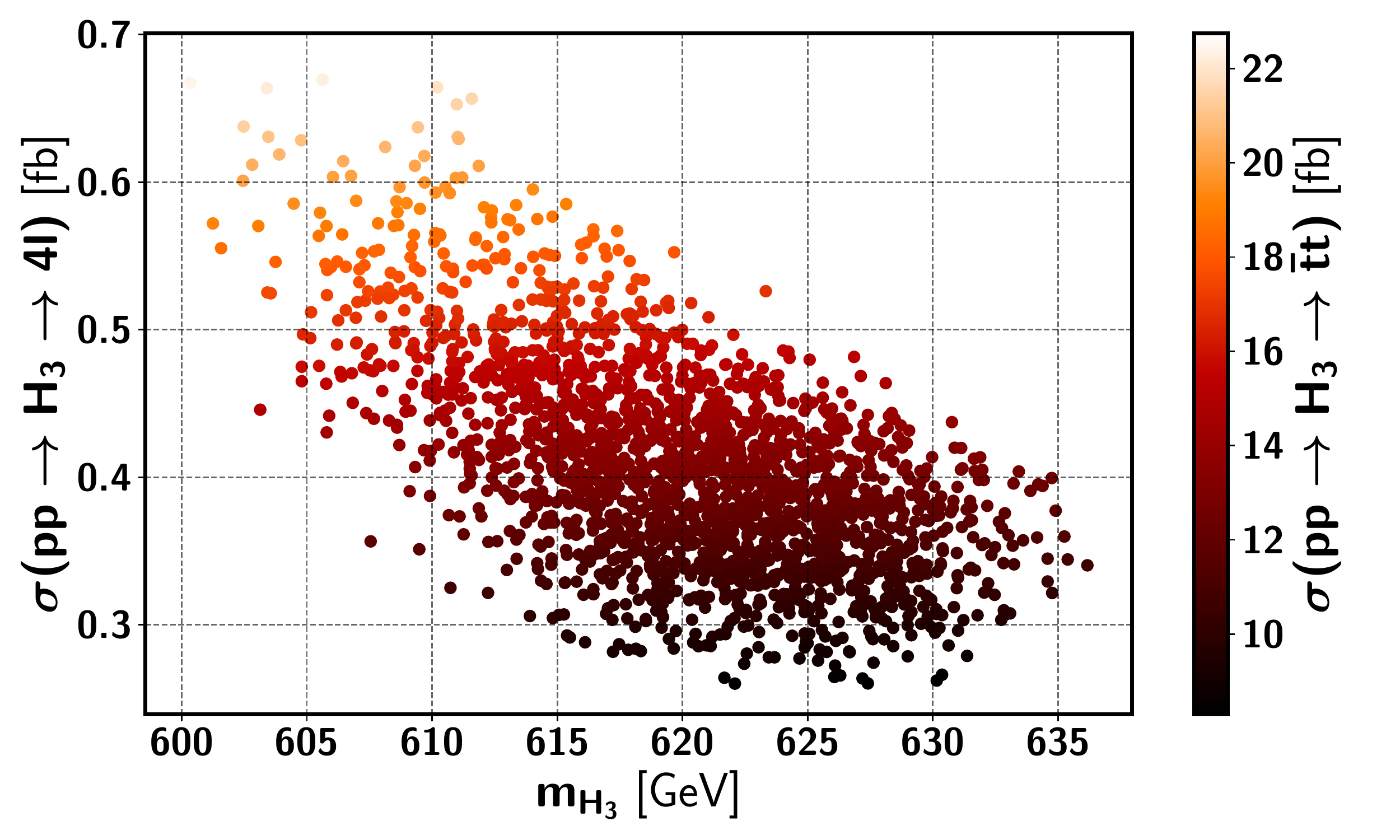}
\end{minipage}

Bottom: $\gamma\gamma$ 

\begin{minipage}{0.49\textwidth}
	\includegraphics[width=\textwidth]{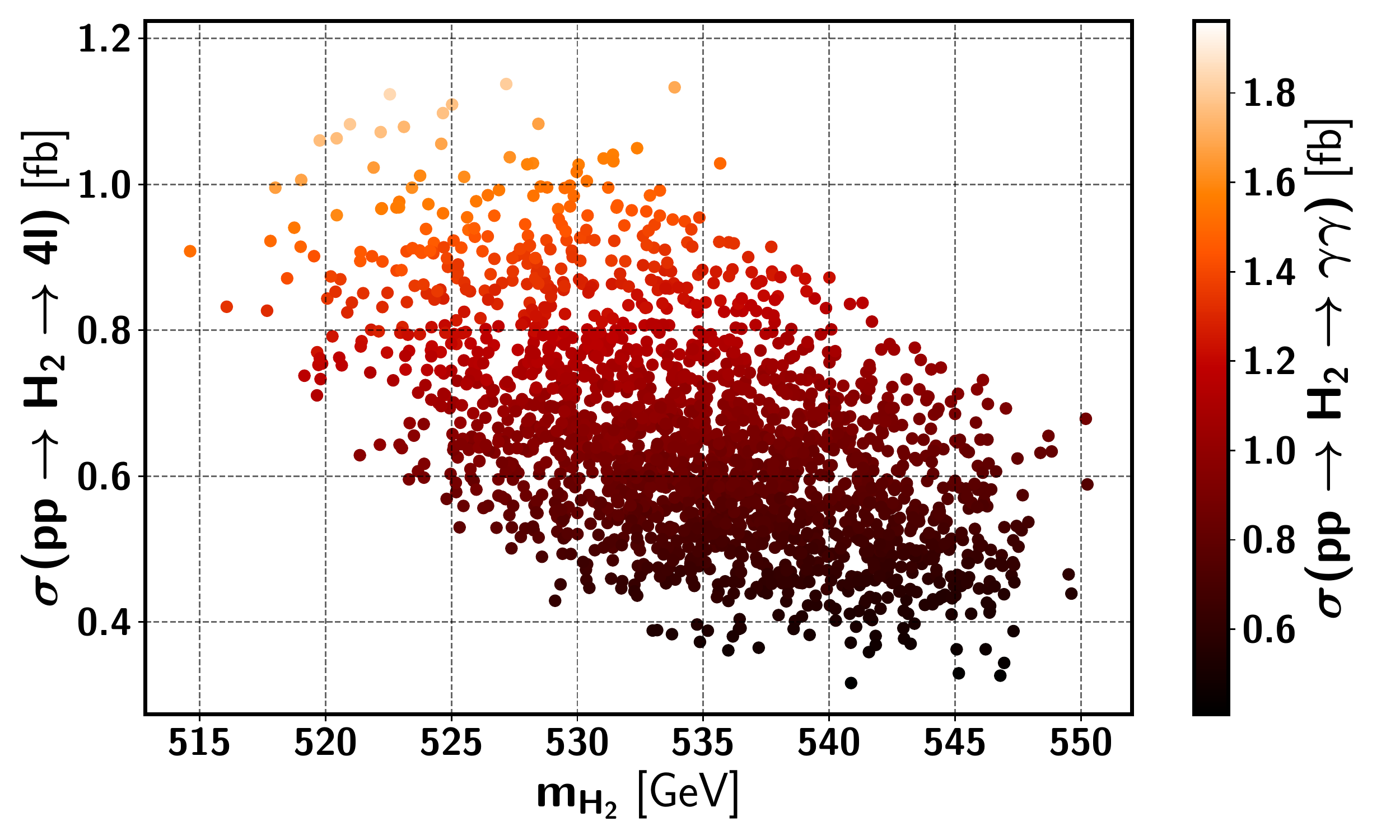}
\end{minipage}
\begin{minipage}{0.49\textwidth}
	\includegraphics[width=\textwidth]{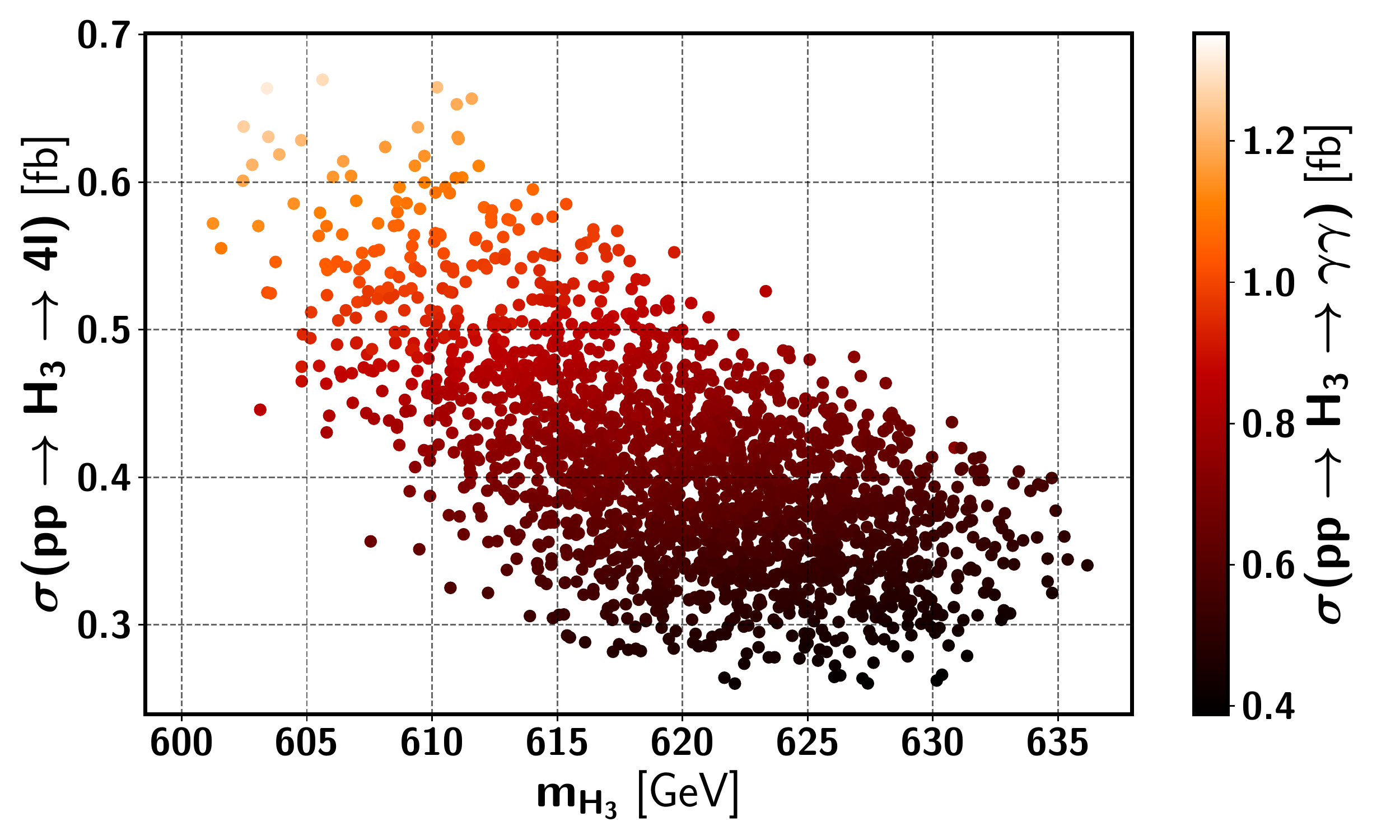}
\end{minipage}

\caption{Results from a very fine-grained parameter space scan within the class of THDMs with CP violation and type I Yukawa structure (cf.\ \cite{Antusch:2020ngh}). Shown points feature masses $m_{H_i}$ and $\sigma_{pp\to H_i \to 4\ell}$, $i=2,3$ within the 90\% Bayesian confidence limit of the best-fit point. The left and right panels are for $H_2$ and $H_3$, respectively, showing total cross sections for the $W^+W^-$ (top), $t\bar t$ (middle) and $\gamma\gamma$ (bottom) final states.}
\label{fig:parameterspacescan}	
\end{figure}

In general, scalars that mix with the Higgs doublet can also decay into other SM particles than $Z$ bosons.
Thus, after finding a good benchmark point to match the $4\ell$ invariant mass spectrum as reported by ATLAS, we explore the possibility of making quantitative predictions for the $H_i$ decays into $t\bar t$, $W^+W^-$ and $\gamma \gamma$.
Therefore, we show in the four panels of fig.~\ref{fig:parameterspacescan} the projections of the total cross sections for the $\sim 2000$ parameter space points from the very fine grained scan over THDM model parameters.
The figure shows the $4\ell$ final state of $H_2$ and $H_3$ into $ZZ$ and the color code of the points denotes the total production cross sections for $W^+W^-$ (upper panels), $t\bar t$ (middle pannels) and $\gamma \gamma$ (lower panels).

\subsection{Discussion}
Our ``best-fit benchmark point'' $P_7$ provides an excellent fit to the excesses of events in the $4\ell$ spectrum with invariant masses between 500~GeV and 700~GeV observed by the ATLAS collaboration. 
The fit prefers $H_2$ and $H_3$ with similar masses and production cross sections for the $4\ell$ final state. Combined with the facts that (i) the THDM adds one CP-even and one CP-odd scalar to the SM field content and (ii) that the CP-odd field does not decay into $ZZ$ this implies that the mass eigenstates must be strongly CP-mixed.
The CP properties of the two scalars could be tested at the HL-LHC via correlations in final states with two tau leptons from the processes $pp \to H_i \to \tau^+ \tau^-$ as discussed in ref.~\cite{Antusch:2020ngh}.

A comment on our statistical analysis is in order at this point. The central aim of this article is {\it not} to establish statistical evidence for the existence of a THDM signal in the $M_{4\ell}$ spectrum. We rather want to point out that {\it if} the here discussed signal were to become statistically significant the THDM with CP violation is a suitable model to explain it.
We thus decided against combining the data sets from ATLAS with a similar analysis by the CMS collaboration~\cite{CMS:2018amk}, which also observes excesses in $4\ell$ final states at high invariant mass. 
Since the CMS analysis uses a smaller data set, including this result would at best increase the statistical significance of our result very slightly.

As mentioned above, our analysis allows us to make quantitative predictions for decays of $H_j,\,j=2,3$ into $t\bar t$.
Our ``best-fit benchmark point'' has cross sections $\sigma_{H_2\to t\bar t} = 22$~fb and $\sigma_{H_3\to t\bar t} =13$~fb. This is smaller than the current uncertainty of the recent measurements of the total production cross section $\sigma_{t\bar t} = 830 \pm 36 {\rm (stat)} \pm 14 {\rm (syst)}$~pb by ATLAS~\cite{ATLAS:2020aln} and $791 \pm 25$~pb by CMS~\cite{CMS:2021vhb}.

Let us now confront our  ``best-fit benchmark point'' with the fact that no enhancement of semi-leptonic final states from $WW$ or $ZZ$ decays has been reported in ref.~\cite{ATLAS:2020fry}.
For our  ``best-fit benchmark point'', the sum of $WW$ and $ZZ$ cross sections yields: 
\begin{equation}
\sum_{i,V} \sigma_{pp\to H_i \to VV} = 392~\rm{fb}\,.
\end{equation}
This cross section is comparable with the $2\sigma$ upper limits on the production cross section from ref.~\cite{ATLAS:2020fry}, which are 470~fb for a mass of $500$~GeV (for the example of a scalar radion).

The small apparent enhancement of the $b\bar b b\bar b = 4b$ final state for invariant masses above 500~GeV as observed in ref.~\cite{ATLAS:2020jgy} could be another indication for the process $pp\to H_i \to ZZ$, $i=2,3$. For the corresponding cross section one would expect that
\begin{equation}
\sigma_{pp\to H_i \to 4b} \geq \left(\frac{{\rm Br}(Z\to b\bar b)}{{\rm Br}(Z\to \ell^+\ell^-)}\right)^2 \epsilon_b^4  \frac{\epsilon_{4b}}{\epsilon_{4\ell}}
\label{eq:Hito4b}
\end{equation}
with the $b$-tagging efficiency $\epsilon_b \simeq 0.7$, and the selection efficiencies $\epsilon_{4b}\sim 0.1$, and where additional $4b$ production could come from $pp \to H_i \to 2 H_1 \to 4b$.
Eq.~\eqref{eq:Hito4b} results in a lower limit of 16.5 additional events in the $4b$ final state, which matches quite well the observed $\sim 20$ events in excess of the background for $M_{4b}\geq 500$~GeV.

Next we comment on the $H_i \to \gamma\gamma$ channel, for which our ``best-fit benchmark point'' has cross sections $\sigma_{H_j\to \gamma\gamma} = 1.2$~fb and $1.05$~fb for $j=2,3$, respectively.
The ATLAS search for resonances in diphoton final states places at $2\sigma$ upper limits on the production cross section of 1.15~fb and 0.83~fb for resonances with masses corresponding to 540~GeV and 631~GeV, respectively \cite{ATLAS:2021uiz}.
This implies that our ``best-fit benchmark point'' is slightly in tension with these limits, but may still be regarded as compatible. We also note that in the current data there indeed exist some upward fluctuations of the observed event counts at 540~GeV and at 680~GeV.
In any case, future analyses of the diphoton spectrum with more data should be able to test this prediction of our  ``best-fit benchmark point''.

Last but not least we consider the impact that CP violation in the scalar sector has for low-energy observables.
As mentioned above, large CP-violating phases are an implication of $H_2$ and $H_3$ having similar signal strengths in the $4\ell$ final state.
Large CP phases imply that the THDM fields give rise to the Electric Dipole Moments (EDM) of SM particles, in particular for the electron, cf.\ e.g.\ refs.~\cite{Abe:2013qla,Chun:2019oix,Altmannshofer:2020shb}. 
In our fine-grained parameter space scan we reject EDM for the electrons that are above the current exclusion limit of $|d_e|<1.1\times10^{-29}$~ecm \cite{Andreev:2018ayy}, and we find that the majority ($\gtrsim 90\%$) of all points has $10^{-30}\leq \frac{|d_e|}{\rm ecm} \leq 1.09 \cdot 10^{-29}$.

\section{Conclusions}
In this paper we have considered a ``double peak'' from a CP violating Two Higgs Doublet Model as explanation for the local excess in four-lepton events with invariant masses above 500 GeV as observed by ATLAS and CMS.
Within a class of THDMs, we used an iterative fitting procedure to search for model parameters that give rise to heavy Higgs masses and signal strengths towards explaining the excess.

The ``best-fit benchmark point'' (called P7) we found this way provides an excellent explanation for the data, with a $\chi^2=5.76$ for 8 bins, predicting events in excess of the SM in the range from 500~GeV to around 700~GeV.
It prefers a broad ``double peak'' in the invariant mass spectrum, with two resonances at 540~GeV and 631~GeV, respectively. 

Interpreted in the context of a THDM, this would be an indication for CP violation in the scalar sector.
The CP mixing is required to be close to maximal due to the comparable production cross section for the two processes $pp\to H_2\to 4\ell$ and $pp\to H_3\to 4\ell$. The CP mixing of $H_2$ and $H_3$ could in principle be tested at the HL-LHC in the future, e.g.\ via the di-tau final state (cf.\ \cite{Antusch:2020ngh}).

Our ``best-fit benchmark point'' predicts additional $t\bar t$, $VV$ and $\gamma\gamma$ production channels with cross sections (summed over $H_2$ and $H_3$) of about 35~fb for $t\bar t$, 199~fb for $WW$ and  2.2~fb for $\gamma\gamma$.
Our results are compatible with present limits, and may be responsible for minor excesses in the $4b$ and $\gamma\gamma$ channels. 
Moreover, the parameter space that leads to an explanation of the observed $4\ell$ spectrum gives rise to electron EDMs that are close to the current experimental bounds, providing an example for a complementary way to test the scenario by low energy experiments.

\subsection*{Acknowledgements}
This work was supported by the Swiss National Science Foundation. A. Hammad is supported from the Basic Science Research Program through the National Research Foundation of Korea Research Grant No. NRF-2021R1A2C4002551. CS was supported by the Cluster of Excellence Precision Physics, Fundamental Interactions, and Structure of Matter (PRISMA+ EXC 2118/1) funded by the German Research Foundation (DFG) within the German Excellence Strategy (Project ID 39083149), and by grant 05H18UMCA1 of the German Federal Ministry for Education and Research (BMBF).

\bibliographystyle{unsrt}

\end{document}